\documentclass[apj]{emulateapj}

\begin{document}
\title{
Interpretation of the Extragalactic Radio Background
}
\author{
M. Seiffert\altaffilmark{1}, 
D. J. Fixsen\altaffilmark{2,3},  
A. Kogut\altaffilmark{2}, 
S. M. Levin\altaffilmark{1},  
M. Limon\altaffilmark{4}, 
P. M. Lubin\altaffilmark{5},
P. Mirel\altaffilmark{2,6},  
J. Singal\altaffilmark{7}, 
T.~Villela\altaffilmark{8},
E. Wollack\altaffilmark{2},
C. A. Wuensche\altaffilmark{8}
}

\altaffiltext{1}{Jet Propulsion Laboratory, 4800 Oak Grove Drive, Pasadena, CA 91109; Michael.D.Seiffert@jpl.nasa.gov}
\altaffiltext{2}{Code 665, Goddard Space Flight Center, Greenbelt, MD 20771}
\altaffiltext{3}{University of Maryland}
\altaffiltext{4}{Columbia Astrophysics Laboratory, 550W 120th St., Mail Code 5247, New York, NY 10027-6902}
\altaffiltext{5}{University of California, Santa Barbara, CA}
\altaffiltext{6}{Wyle Informations Systems}
\altaffiltext{7}{Kavli Institute for Particle Astrophysics and Cosmology, 
SLAC National Accelerator Laboratory, Menlo Park, CA 94025}
\altaffiltext{8}{Instituto Nacional de Pesquisas Espaciais, Divis\~ao de Astrof\'{\i}sica, Caixa Postal 515, 12245-970 -  S\~ao Jos\'e dos Campos, SP, Brazil}

\begin{abstract}
We use absolutely calibrated data between 3 and 90 GHz from the 2006 balloon flight of
the ARCADE~2 instrument, along with previous measurements at other frequencies,
to constrain models of extragalactic emission.
Such emission is a combination of the Cosmic Microwave Background (CMB)
monopole, Galactic foreground emission, the integrated contribution of radio emission from external galaxies,
any spectral distortions present in the CMB, and any other extragalactic source.
After removal of estimates of foreground emission from our own Galaxy, and the estimated
contribution of external galaxies, 
we present fits to a combination of the flat-spectrum CMB
and potential spectral distortions in the CMB. 
We find 2~$\sigma$ upper limits to CMB spectral distortions
of $\mu < 5.8 \times 10^{-5}$ and 
$ |Y_{\mbox{\scriptsize ff}}| < 6.2 \times 10^{-5} $. 
We also find a significant detection of a residual signal beyond that which 
can be explained by the CMB plus the integrated radio emission from 
galaxies estimated from existing surveys. 
After subtraction of an estimate of the contribution of discrete radio sources,
this unexplained signal is consistent with extragalactic
emission in the form of a power law with amplitude $1.06 \pm 0.11$~K at 1~GHz and a spectral index
of $-2.56 \pm 0.04$. 

\end{abstract}

\section{Introduction}
The Cosmic Microwave Background (CMB) is currently our most precise window
on the physics of the early universe. 
Measurements of the frequency spectrum of the CMB can rule out alternative
cosmologies and place limits on physical processes that may distort the spectrum,
including dark matter particle decay and reionization. Departures from a thermal
blackbody spectrum are expected at a small level from a variety of mechanisms.

The {\sl Cosmic Background Explorer} ({\sl COBE}) satellite observed the 
spectrum of the CMB with the Far-Infrared Absolute Spectrophotometer (FIRAS)
instrument \citep{Mather90} at wavelengths between 1 cm and 100~$\mu$m. FIRAS
results reported by \citet{Fixsen96}, \citet{Mather99} and \citet{FixsenMather02} are consistent
with a blackbody spectrum at a temperature of $T_{\mbox{\scriptsize CMB}}=2.725 \pm 0.001$~K.

Absolutely calibrated measurements of the CMB at longer wavelengths (lower 
frequency) than FIRAS have been performed with ground-based and
balloon-borne experiments. Among the most sensitive of these measurements
are those of \citet{johnson87}, \citet{levin92}, \citet{bersanelli94},
\citet{bersanelli95}, \citet{staggs96a}, \citet{staggs96b}, \citet{rag00}, 
\citet{fixsen04}, \citet{singal06}, and \citet{zannoni08}. 

The second generation of the Absolute Radiometer for Cosmology, Astrophysics, and Diffuse Emission (ARCADE~2) 
was conceived as a balloon-borne experiment to improve constraints on spectral
distortions in the CMB, with particular emphasis on the 3 --- 10 GHz frequency range. 
ARCADE~2 uses a unique, clear aperture instrument design with the bulk of the instrument
operating at or near the temperature of the CMB. This minimizes the potential contribution
to instrument systematics from warm, emissive optics. The instrument uses a set of 
microwave feed horns to compare the sky to a large, cryogenic blackbody calibration target.
The results described here are from the second version of the instrument,
described in detail by \citet{Singal08}. 
The sky measurements from the second flight of this instrument are described by \citet{Fixsen08},
which presents a detection of extragalactic emission consistent with a power law plus 
constant CMB temperature.
The model of Galactic emission used in interpreting the ARCADE~2 data is described by \citet{Kogut08}.

In this paper, we use the combination of ARCADE~2 and other data sets 
to present a detection of 3 GHz emission in excess of that expected from the CMB and
existing source counts of radio galaxies. The excess emission amplitude and spectral index
described here differ from the extragalactic emission described by \citet{Fixsen08}.
Here, we are concerned with the residual excess emission that cannot be explained
from known classes of emission, and explicitly correct for the estimated contribution of 
discrete radio galaxies.
We also use the combined data to place limits
on spectral distortions to the CMB, and show that canonical spectral distortions
cannot explain the excess emission.

This paper is organized as follows: Section 2 summarizes the estimates of isotropic,
extragalactic emission at a variety of frequencies that we have used in our analysis.
Section 3 examines the potential contribution of extragalactic point sources and their
potential to affect our conclusions. Section 4 presents our spectral fits to the data and
our limits on spectral distortions of the CMB. Section 5 presents discussion of the results,
including potential explanations for the source of the excess emission.

\section{Results from ARCADE~2 and Other Surveys}

For our analysis, we use the data from the 2006 flight of the ARCADE~2 instrument, 
from FIRAS, and from lower frequency ground-based surveys. FIRAS measures a high-precision
difference between the sky and a calibrated reference target. 
The result is a set of  values with tiny relative errors, and a larger, 1~mK calibration
error common to all the data points.
Table~\ref{table:data} summarizes the remaining data used in our analysis, which includes
ARCADE~2, the 22~MHz survey of \citet{Roger}, the 45~MHz survey of \citet{Maeda},
the 408~MHz survey of \citet{Haslam}, and the 1.42~GHz survey of \citet{ReichReich}. \citet{Kogut08}
describes the process of estimating the Galactic component from each of these data sets. The
data in Table~\ref{table:data} is the resulting
estimate of the residual, extragalactic, isotropic emission.
The ARCADE~2 data in the 3 - 10~GHz range are shown in Figure 1; they lie significantly above
the 2.725~K blackbody CMB determined by FIRAS at higher frequencies.
\begin{figure}[th]
 \begin{center}
	\includegraphics[width=3.5in]{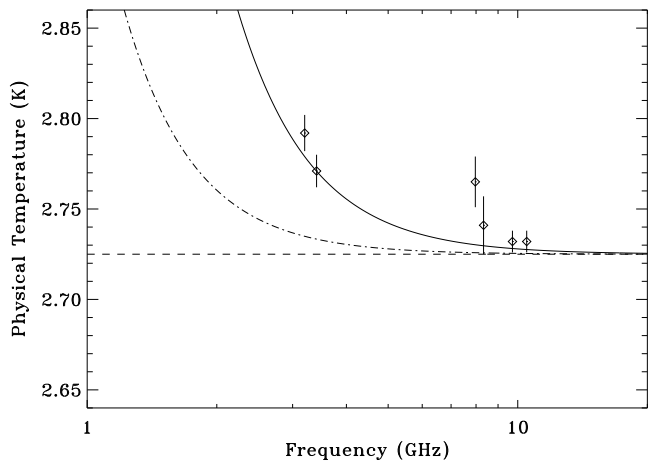}
 \end{center}
\caption{{\small 
Detection of extragalactic radio emission by ARCADE~2 beyond the contribution of discrete radio sources
and the expectation of 2.725~K blackbody radiation. Data points are the ARCADE~2 results from
\citet{Fixsen08}, and have been corrected for Milky Way Galactic emission described by \citet{Kogut08}.  The dashed curve is a constant 2.725~K blackbody, consistent with FIRAS measurements
of the CMB.
The dot dash curve is an estimate of the discrete radio source contribution from \citet{Gervasi08a} model ``Fit1'' added to the 2.725 blackbody.  
The data points lie significantly above this dot dash curve, indicating our detection of unexplained, excess 
emission. 
The solid curve is the best fit of the combined data of Table~1 and FIRAS to a power law plus a constant CMB temperature. 
}}
\end{figure}

\begin{deluxetable*}{ccccccc}
\tablecaption{
Measurements of Extragalactic Radio Emission
\label{table:data}}
\tablehead{
\colhead{Frequency (GHz)} & \colhead{Extragalactic\tablenotemark{a} Temperature (K)} & \colhead{Error (K)}
& \colhead{Radio Sources\tablenotemark{b}} & \colhead{Error\tablenotemark{c}}
& \colhead{Residual Emission\tablenotemark{d}} & \colhead{Error\tablenotemark{e}}
}
\startdata
0.022	&21200\tablenotemark{f}	&5125\tablenotemark{f} &  7090   &  709 & 14100 & 5175 \\
0.045	&4355\tablenotemark{g}		&520\tablenotemark{g} &  1020   &  102 & 3334 & 530 \\
0.408	&16.24\tablenotemark{h}	&3.40\tablenotemark{h} &  2.61   &  0.26 & 13.6 & 3.41\\
1.42		&3.213\tablenotemark{j}	&0.53\tablenotemark{j} &  0.089  &  0.009 & 3.124 & 0.53 \\
3.20&2.792\tablenotemark{k} &0.010\tablenotemark{k} &  0.010  &  0.001 & 2.706 & 0.010 \\
3.41&2.771\tablenotemark{k}&0.009\tablenotemark{k}  &  0.008 	&  $<$ 0.001 & 2.682 & 0.009\\
7.97&2.765\tablenotemark{k}&0.014\tablenotemark{k}  &  $<$ 0.001 & $<$ 0.001 & 2.577 & 0.014\\
8.33&2.741\tablenotemark{k}&0.016\tablenotemark{k}&  $<$ 0.001 & $<$ 0.001 & 2.545 & 0.016\\ 
9.72&2.732\tablenotemark{k}&0.006\tablenotemark{k}&  $<$ 0.001 & $<$ 0.001 & 2.505 & 0.006\\
10.49&2.732\tablenotemark{k}&0.006\tablenotemark{k}&  $<$ 0.001 & $<$ 0.001 & 2.488 & 0.006\\
29.5&2.529\tablenotemark{k}&0.155\tablenotemark{k}&  $<$ 0.001 & $<$ 0.001 & 1.887 & 0.151 \\
31&2.573\tablenotemark{k}&0.076\tablenotemark{k}&  $<$ 0.001 & $<$ 0.001 & 1.900 & 0.074\\
90&2.706\tablenotemark{k}&0.019\tablenotemark{k}&  $<$ 0.001 & $<$ 0.001 & 1.098 & 0.015\\
\enddata
\tablenotetext{a}{This is the monopole temperature with the Milky Way Galactic contribution removed 
as by \citet{Kogut08} }
\tablenotetext{b}{Estimate of extragalactic discrete radio source contribution from \citet{Gervasi08a} model ``Fit1''. Units are K, antenna temperature. }
\tablenotetext{c}{We have adopted a 10\% fractional error for the \citet{Gervasi08a} ``Fit1'' model (see text).}
\tablenotetext{d}{Residual extragalactic emission after subtraction of radio source. Values have been converted to K, antenna temperature.}
\tablenotetext{e}{Error in residual extragalactic emission after subtraction of radio source, K, antenna temperature.}
\tablenotetext{f}{Data from \citet{Roger} corrected for Galactic emission with the model described by \citet{Kogut08}. Units are K, antenna temperature.}
\tablenotetext{g}{Data from \citet{Maeda} corrected for Galactic emission with the model described by \citet{Kogut08}. Units are K, antenna temperature.}
\tablenotetext{h}{Data from \citet{Haslam} corrected for Galactic emission with the model described by \citet{Kogut08}. Units are K, antenna temperature.}
\tablenotetext{j}{Data from \citet{ReichReich} corrected for Galactic emission with the model described by \citet{Kogut08}. Units are K, antenna temperature.}
\tablenotetext{k}{ARCADE~2 \citet{Fixsen08}. Units are K, physical temperature.}
\end{deluxetable*}

In our analysis, we have excluded the 100-200 MHz results of \citet{RogersBowman08}. They
find a minimum diffuse background of 237~K at 150 MHz, but their work does
not provide an independent estimate of the Galactic contribution. 
We can, however, check for consistency by using the Galactic model described by \citet{Kogut08} extrapolated
to 150 MHz, where we find an approximately 60~K Galactic contribution to the diffuse
background in the region of high Galactic latitude. Applying this correction, we find
both the emission amplitude and spectral index are consistent with the fits we present
in section 4.
%
%We have similarly excluded the results reported by \citet{Bridle67} at 178 MHz ($37 \pm 8$K) for
%an isotropic component. Here the agreement with the rest of the data is not as good, and
%is in mild disagreement with \citet{RogersBowman08}.

We have not included a number of other measurements, including the rocket-borne measurements
of \citet{Gush90} and the ground-based and balloon-borne measurements cited earlier. The
size of the uncertainties quoted in these measurements results in no significant 
contribution to the constraints on our model fits.

\section{Contribution of Sources} 
The set of measurements in Table~1 do not have sufficient angular resolution
to reject discrete radio sources. Instead, we must
estimate the contribution of these sources through one of two ways: direct
radio surveys designed to detect such sources, or measurements of
the far-IR background which can trace the integrated emission of such sources through 
the correlation of the far-IR and radio emission. We examine these two methods in turn.

\subsection{Expectation from source counts}

The sky brightness temperature contributed by discrete sources can be composed as the sum of 
two parts: the source population that has been characterized by existing surveys and the
contribution of sources below the flux limit of existing surveys. We write this as:
\begin{equation}
T = T(S > S_{\mbox{\scriptsize limit}}) + \frac{\lambda^2}{2 k_{\mbox{\scriptsize B}}} \int_{S_{\mbox{\scriptsize min}}}^{S_{\mbox{\scriptsize limit}}}
	\frac{dN}{dS} S dS,
\end{equation}
where $T(S > S_{\mbox{\scriptsize limit}})$ is contribution from sources with a flux $S$ greater than the 
survey limit $S_{\mbox{\scriptsize limit}}$. The wavelength of observation is $\lambda$ and  $k_{\mbox{\scriptsize B}}$
is the Boltzmann constant. We characterize sources below the survey limit with their differential
number counts, ${dN}/{dS}$, and assume that there is a lower limit cutoff to the source population
at a flux of $S_{\mbox{\scriptsize min}}$.  Radio source count surveys reveal a
faint source population with differential number counts proportional to a power law
\begin{equation}
\frac{dN}{dS} = S^{-\gamma}.
\end{equation}
An index $\gamma$ of 2.5 corresponds to a static, Euclidean universe with uniform filling of
sources, whereas faint radio surveys find $\gamma$ in the range of $2.0$ to $2.3$. Such source counts can not
extend to arbitrarily low fluxes, or the total contribution would diverge. 
A realistic distribution of sources, of course, would not have a sharp cutoff at 
$S_{\mbox{\scriptsize min}}$. In practice, we can characterize $S_{\mbox{\scriptsize min}}$
as the flux below which the index $\gamma$ falls below 2, 
as there will be negligible additional contribution to the integral below this limit. 

Deep surveys of radio sources have been performed at a number of frequencies. 
Particularly useful are the surveys at 1.4 and 8.4 GHz with the Very Large Array (VLA).
\citet{Fomalont02} reports the results of an 8.4 GHz survey with a survey limit of
$7.5 \mu$Jy and finds a faint-end index to the number counts of $\gamma = 2.11 \pm 0.13$.
\citet{Windhorst93} argue that the sources in the nJy flux range are dominated by ordinary spiral galaxies,
which produces a natural lower limit of $30$~nJy; below this limit there are insufficient
galaxies. 

\begin{figure}[th]
 \begin{center}
	\includegraphics[width=3.5in]{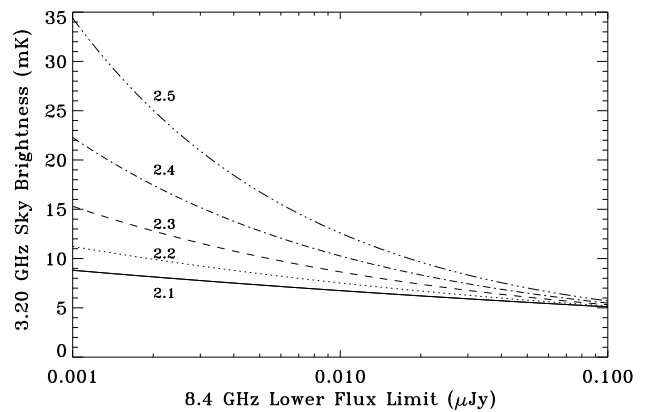}
 \end{center}
\caption{{\small 
Estimated contribution of extragalactic source counts to the sky brightness at 3.20 GHz, versus
the assumed faint end cutoff of 8.4 GHz source counts. The sky brightness is first calculated
at 8.4 GHz from the sum of the existing source population greater than 7.5 $\mu$Jy (Fomalont et
al 2002) and the contribution of fainter sources assuming a continuation of the differential
number counts with a power law index of 2.1 (solid curve), 2.2 (dotted curve), 2.3 (dashed curve),
2.4 (dash dot curve), and 2.5 (dash dot dot curve). The contribution is scaled to the ARCADE 3.20 
GHz channel with a frequency index of -2.75, as is typical of sources in faint radio surveys.
The measured index of differential number counts at the faint end of the 8.4 GHz survey is $\gamma = 2.11 \pm 0.13$.
}}
\end{figure}

Figure~2 shows a range of estimates for the contribution of discrete sources to
the ARCADE~2 3.20 GHz measurement. We have calculated the expected contribution
to the extragalactic sky temperature using the results of \citet{Fomalont02} and
varying the faint end index, and plotting as a function of $S_{\mbox{\scriptsize min}}$.
We have scaled the temperature from 8.4 to 3.2~GHz using a frequency 
spectral index of -2.75. This spectral index is characteristic of starburst and normal
spiral galaxies with synchrotron emission, though at 8.4~GHz there could be some
contribution from flat spectrum sources such as AGN. The sources fainter than 35~$\mu$Jy described
by \citet{Fomalont02} have a spectral index distribution that peaks at -2.75.  
From this analysis, we conclude that the contribution is likely 
in the range of 5 to 10 mK at 3.20 GHz. 

It is interesting to consider to what extremes one would need to take the number counts
in this analysis to account for the measured 58~mK excess at 3.20 GHz. Using an index of
$\gamma = 2.5$ and extrapolating the number counts from \citet{Fomalont02}, we find
that we would need to extend the lower flux limit to 0.3~nJy, which would result
in a source density of $\sim 8 \times 10^5$ per square arcmin. This extreme scenario
is not a plausible distribution of ordinary galaxies. 
We also note that the normalization of the differential number counts in the 8.4~GHz survey
is sufficiently accurate to not contribute a significant source of error to this analysis.

We have focused on the 8.4~GHz survey results, as these measurements are amenable
to extrapolation to fainter fluxes and to the ARCADE~2 frequencies. Our results, however,
do not depend on the results from a radio survey at one frequency.
\citet{Gervasi08a} describe a more comprehensive analysis of the potential contribution of 
unresolved extragalactic radio sources by examining data from a wide variety of
radio surveys, and fitting an empirical, two-population model to the survey data at
each frequency. Their combined result for extragalactic radio source brightness versus
frequency is described by a single power law model, ``Fit1'', 
with amplitude 0.88~K at 0.61~GHz, and a power law index of -2.707. This model is
also shown in Figure~1, indicating again that this contribution is insufficient to
explain the ARCADE~2 results. The brightness temperature
values of this model, evaluated at the frequencies used in our analysis, appear in Table~1.
Because the sample size and brightness limits of the radio surveys vary with frequency, the
error in their estimate also varies with frequency. These frequencies are not identical to
the frequencies measured by ARCADE~2. For Table~1, we adopt a fractional error estimate
of 10\% independent of frequency, which we believe is a conservative estimate of the error in
the \citet{Gervasi08a} modeling. These estimates for the contribution of extragalactic sources
are consistent with the analysis described above.

It also is interesting to consider whether the existing surveys have missed significant
flux from the sources. The low frequency faint source observations are primarily interferometric
and have the possibility of overresolving the source and missing flux in extended low surface 
brightness emission. \citet{HenkelPartridge05} consider the evidence for this and conclude
that 20\% may be an upper limit to this effect for mJy flux levels at 8.5 GHz. 
\citet{Fomalont06} suggests
that at 1.4 GHz only a few percent of sources are larger than 4 arc seconds and that
other reports of a larger figure in other surveys are actually confusion of multiple
disparate sources. 
\citet{Garrett00} compare their 1.4 GHz survey conducted using the Westerbork Synthesis
Radio Telescope and its larger, 15 arcsec effective beam with previous VLA measurements of
the same region with a similar noise level. Of a total of 85 sources in their survey,
they find 22 not apparent in the previous VLA survey. Some of these 22 likely 
correspond to the combined flux of multiple sources that were resolved by the VLA measurements.
At least 2 sources, however, appear to be relatively nearby discrete sources with emission
from a large enough region to have been resolved out by the VLA measurements. 
We conclude from these studies that it seems unlikely that sufficient flux has been missed
in surveys of known objects to explain our residual emission. 

Another method to examine the possibility of extended low surface brightness emission in
extragalactic sources is radio observations of the halos of nearby edge on spirals.
\citet{Irwin99} and \citet{Irwin00} report results of VLA surveys for radio emission from nearby edge-on spirals. 
These studies can elucidate the connection
between the star formation processes that drive the far-IR background, and the supernova processes
that drive radio emission, but do not provide evidence that large amounts of radio flux
are missed in surveys of more distant sources.

\subsection{Connection with Far-IR Background}

The cosmic far-IR background has been detected at a level of approximately 
$10-20 \,\mbox{nW}\, \mbox{m}^{-2} \, \mbox{sr}^{-1}$ with FIRAS and DIRBE
\citep{Puget96, Fixsen98, Hauser98}.
We can use the universal radio to far-IR correlation in star-forming galaxies \citep{Condon92}
to estimate the expected extragalactic radio background that can be attributed to galaxies
contributing to the cosmic far-IR background. 
%The intrinsic dispersion of the
%radio to far-IR SED is discussed in Carilli and Yun (1999), Yun and Carilli (2002)
%Carilli and Yun (1999) calculate a spectral index between 350 GHz (850 $\mu$m) and 1.4 GHz as a function of 
%redshift using models from Condon (1992), with a typical value of 0.5 near redshift 2.
\citet{HaarsmaPartridge00} and \citet{DwekBarker02} specifically address this prospect.
The conclusion
of these studies is that the far-IR measurements are consistent with the existing
surveys of radio galaxies described earlier. For example, the radio brightness temperature of 18~K at 178~MHz
predicted by
\citet{DwekBarker02} is within 1~$\sigma$ of the radio galaxy contribution modeled by
\citet{Gervasi08a}. This emission is insufficient to account for the excess
detected by ARCADE~2.

There is no obvious way around this limit by considering departures from the far-IR radio
correlation associated with faintness or redshift.
\citet{GarnAlexander08} stack IR-selected galaxies and data from faint radio surveys 
and find there is no
evidence for a change in the far-IR to radio correlation with fainter galaxies. Similarly,
there does not appear to be evidence for a change in the far-IR to radio correlation with redshift
\citep{Chapman05,Frayer06}.

There is some room for breaking the far-IR to radio correlation; the physical processes of
far-IR emission from dust heated by star formation and radio emission driven by supernovae are 
related but differ in timescale \citep[see, e.g.,][]{Murphy08}. There does not appear, however, to be a case for a sufficient difference
in timescale to account for our measurements in known populations of galaxies.

\section{Fit of Model Spectrum}
Here we fit a model spectrum 
%1) a CMB baseline temperature, 2) a frequency-dependent power law, and 3) possible CMB spectral distortions, 
to the combined data set of the residual emission column of Table~1 and the higher frequency results from FIRAS. 
The form of the fitting function is:
\begin{equation}
T(\nu) = T_0 + A \left({\nu / \mbox{1 GHz}}\right)^{\beta} + \Delta T(\nu),
\end{equation}
where $T_0$ is the CMB baseline temperature, $A$ is the power law amplitude at 1~GHz, $\beta$ is the power law
index, $\nu$ is frequency, and $\Delta T(\nu)$ is a CMB spectral distortion. Data are converted (where necessary) to 
units of antenna temperature before fitting using:
\begin{equation}
T_{\mbox{\tiny Ant}} = \frac{h \nu / k}{e^{h \nu / k T_{\mbox{\tiny Phys}}} -1},
\end{equation}
where $h$ is Planck's constant, $k$ is Boltzmann's constant, $T_{\mbox{\tiny Ant}}$ is antenna
temperature, and $T_{\mbox{\tiny Phys}}$ is the physical temperature.
We use a 
Levenberg-Marquardt non-linear least-squares minimization for the fitting procedure 
\citep{Marquardt}.  We have allowed for the 1~mK relative calibration uncertainty between the FIRAS temperature 
scale and the other measurements in determining the size of the errors on the fit parameters. 
This was accomplished by adding or subtracting 1~mK to the FIRAS measurements and noting
the change in the parameter value. This change was then added in quadrature to
the error derived without allowing the relative calibration error. In most cases this
makes a only very small difference. The exception is $T_{\mbox{\scriptsize CMB}}$, where we recover the
expected 1~mK error. The parameters for the fits described in this section are summarized
in Table~2.

\subsection{Power-Law Plus CMB}
Figure~1 shows that there is clear excess emission detected by ARCADE~2 in the 3~GHz
channels compared to what is expected from the CMB plus the contribution of extragalactic
radio sources. The unexplained residual emission is consistent
with a power law with amplitude $1.06 \pm 0.11$~K at 1~GHz, with a spectral index
of $\beta = -2.56 \pm 0.04$.  

We have also experimented with inflating the assumed errors
for the removal of extragalactic discrete sources.  As noted in Section~3.1, we
have assumed a fractional error of 10\%, independent of frequency, for the 
discrete source contribution model. Inflating this error to 50\% fractional
error makes much less than a 1~$\sigma$ change in the values of our power law
fit parameters above, and only a small ($\sim 10$\%) increase in the quoted
errors in the fit parameters. This is because the error in the removal of
the discrete sources is smaller than the other errors in the low frequency
measurements, as can be seen by inspection of Table~1.

\subsection{Free-Free Distortions}

Free-free distortions to the CMB spectrum can arise from energy released at
lower redshifts 
\citep{BartlettStebbins91, GnedinOstriker97, Oh99}, and can be characterized by
\begin{equation}
\Delta T_{\mbox{\scriptsize ff}}(\nu)= T_0 \frac{Y_{\mbox{\scriptsize ff}}}{x^2}, 
\end{equation}
where $Y_{\mbox{\scriptsize ff}}$ is the optical depth to free-free emission, 
$T_0$ is the undistorted CMB temperature, $x$ is the dimensionless frequency $h\nu/kT_0$,
and $\Delta T$ is apparent temperature distortion.

A lower limit can be placed on the optical depth to free-free emission 
from late time effects of $Y_{\mbox{\scriptsize ff}} > 8 \times 10^{-8}$
\citep{HaimanLoeb97}. The current upper limit of $Y_{\mbox{\scriptsize ff}} < 1.9 \times 10^{-5}$
comes from combining data from FIRAS and previous ground-based CMB spectrum measurements 
\citep{bersanelli94}.

We have performed a four component fit to the data to assess whether there
is evidence to support a free-free spectral distortion to the CMB, compared to the
three parameter fit described in the previous section.
The four fit components consist of a constant CMB temperature, a power law amplitude,
a power law index, and a free-free amplitude.
The fit parameters are presented in Table~2. 
The addition of the free-free amplitude does not improve the reduced $\chi^2$ of the fit compared
to the fit presented in Figure~1 and is therefore not justified by the data.

We have also examined a two parameter fit consisting of a constant CMB component
and free-free distortion component; the parameters are shown in Table~2.
This fit produces a significantly worse reduced
$\chi^2$, and is therefore not consistent with the source of the unexplained
emission. 

The $2 \sigma$ limits on the free-free amplitude at 1 GHz derived from our four parameter fit are
$ -0.44 \mbox{K} < \Delta {T_{Y_{\mbox{\tiny ff}}}} < 0.52 \mbox{K} $. 
This corresponds to an upper limit on the free-free optical depth of
\begin{equation}
 |Y_{\mbox{\scriptsize ff}}| < 6.2 \times 10^{-5}
\end{equation}
This limit is less constraining than those reported by \citet{bersanelli94} and
\citet{Gervasi08b}. This is the result of the additional degrees of freedom allowed
by our four parameter fit. As described above, the four parameter fit is a better
description of the data than the CMB plus free-free distortion fits performed
by \citet{bersanelli94} and
\citet{Gervasi08b}. We conclude that the tighter constraints offered by those studies
are likely too optimistic.

\begin{figure}[th]
 \begin{center}
    \includegraphics[width=3.5in]{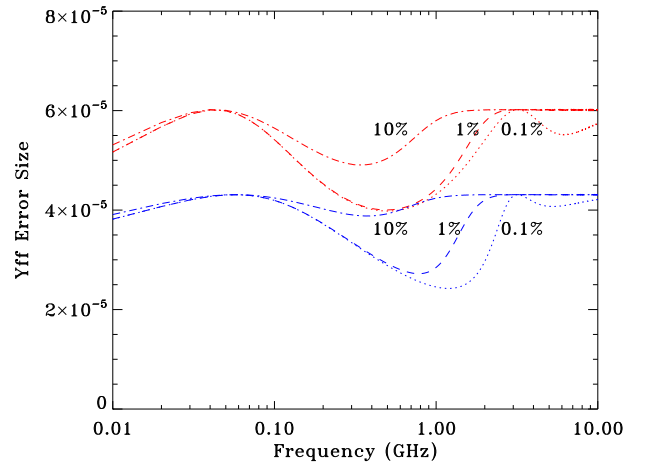}
  \end{center}
\caption{{\small 
Improvement of limit on $Y_{\mbox{\scriptsize ff}}$ obtainable with future measurements. 
Shown are the sizes of the 2~$\sigma$ errors 
on $Y_{\mbox{\scriptsize ff}}$ (red curves) that would result from an additional measurement
at a frequency shown on the x-axis, with a precision of 0.1\% (dotted curve),
1\% (dashed curve), or 10\% (dot dashed curve). The blue curves repeat this
test under the assumption that there are additional measurements at 3, 5, 8, 10, 30, and 90 GHz
with 0.002 K precision, as would result from a future flight of ARCADE~2. The additional measurement is assumed to be centered
on the value of the existing fit. }}
\label{fig:whatif}
\end{figure}

It is interesting to consider how future measurements might improve the 
$Y_{\mbox{\scriptsize ff}}$ limit.
We have used the existing fits and asked how much tighter the limit becomes
if an additional measurement of some fixed fractional accuracy is added to the
data set. We have run this test as a function of frequency; the frequency range
with the greatest effect is between
0.3 and 3.0 GHz, where a factor of several tighter constraint is
potentially achievable. The results are shown in 
Figure~\ref{fig:whatif}, where we have plotted the size of the 2~$\sigma$ errors on
$Y_{\mbox{\scriptsize ff}}$ as a function of the frequency of the additional measurement.  
We have assumed that the
measurement point falls precisely on the existing fit. Better fractional
error results in tighter constraints on $Y_{\mbox{\scriptsize ff}}$. We have also considered
the impact of repeating this test with additional measurements at
3, 5, 8, 10, 30, and 90 GHz with 0.002 K precision, as might be obtained with a future flight
of ARCADE. Such a measurement would enhance the power
of future lower frequency measurements to constrain $Y_{\mbox{\scriptsize ff}}$. 

\begin{deluxetable*}{lcccc}
\tablecaption{
Spectral Fits to Combined ARCADE~2, FIRAS, Radio Survey Data
\label{table:fitresults}}
\tablehead{
\colhead{Parameter} & \colhead{Power Law Fit} & \colhead{Power Law + $Y_{\mbox{\scriptsize ff}}$}
& \colhead{$Y_{\mbox{\scriptsize ff}}$ Only} & \colhead{Power Law + $\mu$ Distortion}
}
\startdata
$T_{\mbox{\scriptsize CMB}}$ \tablenotemark{a} & $2.725 \pm 0.001$ & $2.725 \pm 0.001$  & $2.725 \pm 0.001$ & $2.725 \pm 0.001$ \\
Power Law Amplitude \tablenotemark{b} & $1.06 \pm 0.11$ & $1.00 \pm 0.37$ & \nodata & $1.05 \pm 0.11$ \\
Power Law Index			& $-2.56 \pm 0.04$ & $-2.58 \pm 0.11$ & \nodata & $-2.56 \pm .05$ \\
Free-Free Amplitude \tablenotemark{c} & \nodata & $0.04 \pm 0.24$ & $0.54 \pm 0.07$ & \nodata \\
$\mu$ Amplitude	& \nodata & \nodata & \nodata & $(-0.73 \pm 3.3) \times 10^{-5}$ \\
Degrees of Freedom & 53 & 52 & 54 & 52\\
$\chi^2$ & 60.8 & 60.8 & 107.1 & 60.8\\
Reduced $\chi^2$ & 1.15 & 1.17 & 1.98 & 1.17 \\
\enddata
\tablenotetext{a}{The best fit thermodynamic temperature of the CMB in K.}
\tablenotetext{b}{The fit amplitude of a power law component in K (antenna temperature) at 1 GHz.}
\tablenotetext{c}{The fit amplitude of $\Delta {T_{Y_{\mbox{\tiny ff}}}}$ in K (antenna temperature) at 1 GHz.}

\end{deluxetable*}

\subsection{Chemical Potential ($\mu$) Distortions}

As we have discussed, energy release early in the universe 
can distort the spectrum of the CMB. Energy injection 
after a redshift of $\sim 10^6$  no longer results in a planckian spectrum, but
instead forms a spectrum with a Bose-Einstein
photon occupation number \citep{SZ70}:
\begin{equation}
\eta(x) = \frac{1}{e^{x + \mu(x)} -1},
\end{equation}
where $x \equiv h\nu / k T$ is the dimensionless frequency, and 
$\mu(x)$ is a frequency-dependent chemical potential. A series of 
papers \citep{DaneseDezotti80, Burigana91a, Burigana91b, Burigana95} has
investigated in detail the shape of the resulting spectral distortions
after inclusion of free-free processes which act at the low frequency end.
The shape of the distortion depends on the range of redshift over which such
energy injection takes place.
We have used their analytic description of such distortions to the CMB
to provide a functional form to fit the data in Table~1. A necessary ingredient
is the baryon density, for which we adopt a value of $\Omega_b h^2 = 0.02265$,
from the 5-year WMAP data, including constraints from Baryon Acoustic Oscillation and
Supernova measurements as described by \citet{Hinshaw08}.

\begin{figure*}[th]
 \begin{center}
    \includegraphics[width=6.5in]{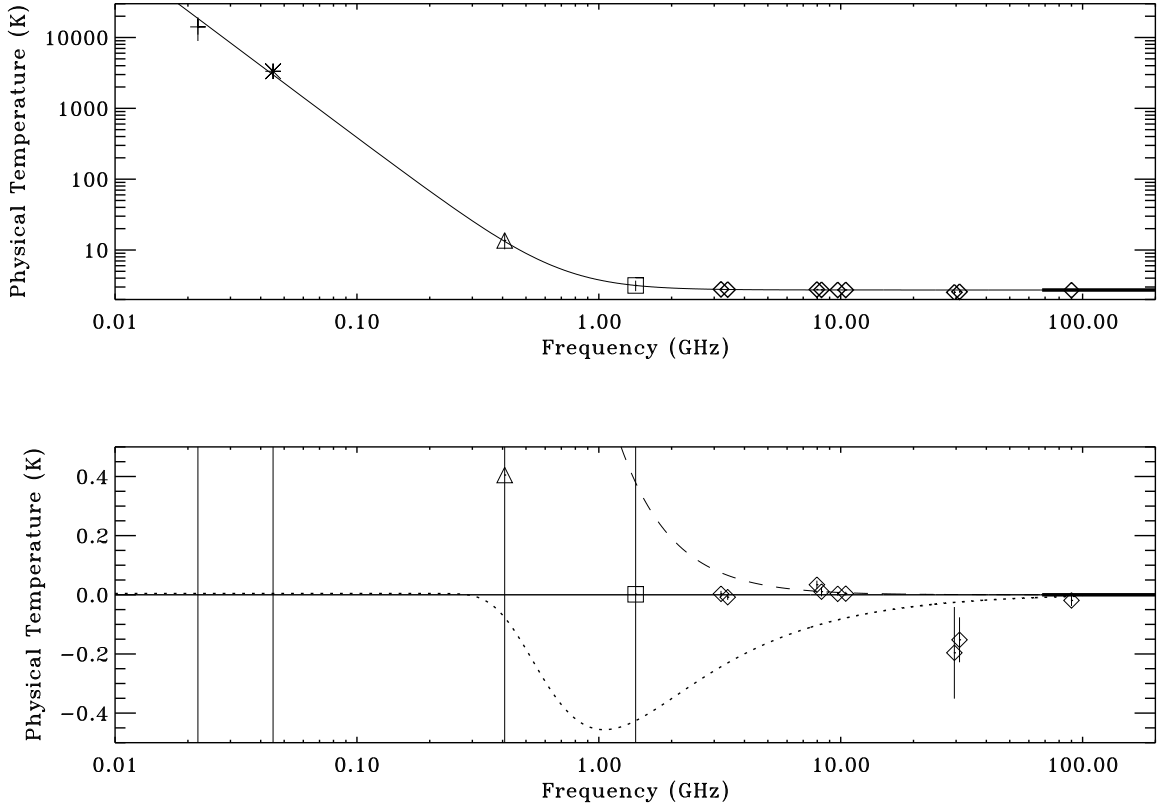}
  \end{center}
  \label{fig:mufit}
\caption{{\small 
Fit of ARCADE~2 data, FIRAS data, and data from low frequency radio surveys. The upper
plot shows (solid line) a fit with three components: a frequency independent CMB contribution,
a power law amplitude, and a power law index.
The lower plot shows the fit residuals. The dotted line shows the expected shape of a $\mu$ distortion. The amplitude of the plotted distortion is
100 times the 2$\sigma$ upper limit allowed by the fit. The dashed line shows the 
shape of a $Y_{\mbox{\scriptsize ff}}$ distortion with an amplitude equal to the 2$\sigma$ upper limit.
The addition of either a $\mu$ distortion or a $Y_{\mbox{\scriptsize ff}}$ distortion 
as a free parameter is not supported by the data. Data points
are from \citet{Roger} (cross), \citet{Maeda} (asterisk), \citet{Haslam} (triangle), \citet{ReichReich} (square), ARCADE~2 (diamonds), and FIRAS (heavy line), corrected for Galactic emission and 
an estimate of extragalactic radio sources, as shown in Table~1.
}}
\label{fig:mu_fit}
\end{figure*}

In addition to the $\mu$ distortion, we have included a power law amplitude and index to the 
fit parameters. Figure~4 shows the result of the fit as well as the residuals. 
The addition of the $\mu$ distortion as a free parameter does not improve the reduced $\chi^2$.
The 2$\sigma$ upper limit on $\mu$ is 
\begin{equation}
\mu < 5.8 \times 10^{-5}.
\end{equation}

This limit is an improvement on the previous limit 
of $9 \times 10^{-5}$
set using FIRAS \citep{Fixsen96}. Although it is numerically similar to the limit reported recently by \citet{Gervasi08b}, we believe it to be a more robust limit because we have allowed additional
free parameters in our fit. 

\section{Discussion}

We have presented evidence for isotropic radio emission detected by ARCADE~2
beyond what can be explained by Galactic emission 
and the unresolved emission from the known population of discrete sources.
The excess emission is consistent with a power law, with an index of $-2.56$, which
is significantly flatter that what might be expected from an unidentified population
of faint, diffuse, steep spectrum (index~$\sim -2.7$) radio sources associated with star-forming galaxies.
We have also examined and placed limits on two classes of spectral distortions
to the CMB. Such distortions are not supported by the data and cannot explain
the excess emission, as is illustrated graphically in Figure 4.
%
%One can extrapolate the power law residual emission to the WMAP 23~GHz channel.
%The result is 350~$\mu$K. 
%The emission must therefore be relative uniform (or be contributed by a large
%number of discrete sources), to escape detection by WMAP, which unfortunately, 
%has insufficient absolute calibration to detect a small uniform signal. 

It is possible to imagine that an unknown population of discrete sources exist
below the flux limit of existing surveys. We argue earlier that these cannot be 
a simple extension of the source counts of star-forming galaxies. 
As a toy model, we consider a population of sources distributed with a delta function in 
flux a factor of 10 fainter than the 8.4 GHz survey limit of \citet{Fomalont02}.
At a flux of 0.75~$\mu$Jy, it would take over 1100 such sources per square
arcmin to produce the unexplained emission we see at 3.20~GHz, assuming a frequency
index of $-2.56$. This source density is more than two orders of magnitude higher than 
expected from extrapolation to the same flux limit of the known source population. It
is, however, only modestly greater than the surface density of objects revealed in the
faintest optical surveys, e.g., the Hubble Ultra Deep Field \citep{Beckwith06}.
The unexplained emission might result from an early population of non-thermal 
emission from low-luminosity AGN; such a source would evade the constraint implied by the far-IR measurements.

We believe our result to be robust to several potential sources of error.
Underestimating the level of Galactic emission is a potential contaminant. As described
by \citet{Kogut08}, however, the expected contribution of the area around the North Galactic
Pole is only $\sim~0.5$~K at 1~GHz, with relatively tight errors. Further, the Galactic emission is estimated with two independent methods along several lines of sight. 

%Relativistic electrons trapped in the Earth's magnetic field emit synchrotron radiation \citep{DyceNakada59}.  %This emission is polarized and anisotropic, however, with intensity peaking near the Earth's magnetic equator %and falling to zero at the poles.  Further, it is measured to be less than 3 K at 30 MHz and expected to be %much less than 1 mK at 3 GHz \citep{Ochs63, PetersonHower63}. It therefore is unlikely to be the source
%of the unexplained emission.
%
Correcting for instrumental systematic errors in measurements such as ARCADE~2 is always a
primary concern.  We emphasize that we detect unexplained emission at 3~GHz with the ARCADE~2 data,
but the result is also independently detected by a combination
of low frequency data and FIRAS. The unexplained emission is detected
above  3$\sigma$ with any two of the three data subsets: 1) FIRAS and low frequency radio data,  
2) ARCADE~2 and low frequency radio data, and 3) ARCADE~2 and FIRAS. 
The result is therefore robust to problems in any one measurement.
%The power of the result comes from the smallness of the error bars
%at 3 GHz combined with the lever arm from the large frequency range of the data in Table~1.

We conclude that the three most important sources of error, 
Galactic emission, instrumental systematic errors, and radio emission from the faint end of
the distribution of known sources, are unlikely to be sufficient to explain the 
excess emission presented here.

It is a pleasure to thank the staff of the Columbia Scientific Balloon
Facility for their support of ARCADE 2.
We thank the undergraduate students whose work helped make
ARCADE 2 possible: Adam Bushmaker, Jane Cornett,
Sarah Fixsen, Luke Lowe, and Alexander Rischard. 
ARCADE 2 was supported by the National Aeronautics and Space
Administration suborbital program. 
T.V. acknowledges support from CNPq grants 466184/00-0, 305219-2004-9, and 303637/2007-2-FA, and the technical support from Luiz Reitano.
C.A.W. acknowledges support from CNPq grant 307433/2004-8-FA.
The research described in this paper
was performed in part at the Jet Propulsion Laboratory,
California Institute of Technology, under a contract with
the National Aeronautics and Space Administration.

\end{document}